\documentclass[pdflatex,aps,twocolumn,superscriptaddress,prl,fleqn,showpacs,nofootinbib,preprintnumbers]{revtex4-1}
\usepackage{amssymb,amsmath,epsfig,color}
\usepackage{times,txfonts}
\usepackage{float} 
\usepackage[caption=false]{subfig} 
\usepackage{ulem}
\usepackage{color}

 \usepackage{ifpdf}
 \ifpdf
 \usepackage[pdftex]{hyperref}
 \else
 \usepackage[hypertex]{hyperref}
 \fi

 \hypersetup{
   pdftitle={},%
   pdfauthor={},%
   pdfsubject={},%
   pdfkeywords={},%
   pdfstartview={},%
   bookmarksopen=true, breaklinks=true, debug=true, %
   colorlinks=true, linkcolor=red, citecolor=blue, urlcolor=blue
 }

\renewcommand{\d}{\mathrm{d}}
\newcommand{\sT}{{\scriptscriptstyle T}}

\renewcommand{\bm}[1]{\mbox{\boldmath $#1$}}

\newcommand{\kTsqav}{\langle p_\sT^2 \rangle}

\newcommand{\mc}[1]{\mathcal{#1}}
\newcommand{\kT}{\bm{k}_{\sT}}
\newcommand{\koneT}{\bm{k}_{1\sT}}
\newcommand{\ktwoT}{\bm{k}_{2\sT}}
\newcommand{\qT}{\bm{q}_\sT}
\newcommand{\ie}{{\it i.e.}}
\newcommand{\eg}{{\it e.g.}}

\newcommand{\cf}[1]{{Fig.~\ref{#1}}}
\newcommand{\ce}[1]{{Eq.~\ref{#1}}}

\def\slash#1{\setbox0=\hbox{$#1$}               
        \dimen0=\wd0                            
        \setbox1=\hbox{/} \dimen1=\wd1          
        \ifdim\dimen0>\dimen1                   
        \rlap{\hbox to \dimen0{\hfil/\hfil}}    
        #1                                      
        \else              
        \rlap{\hbox to \dimen1{\hfil$#1$\hfil}} 
        /                                       
        \fi}                                    %

\begin{document}

\title{Accessing the Transverse Dynamics and Polarization of Gluons inside the Proton at the LHC}

\preprint{NIKHEF-2013-040}

\author{Wilco J.~den Dunnen}
\email{wilco.den-dunnen@uni-tuebingen.de}
\affiliation{Institute for Theoretical Physics, Universit\"{a}t T\"{u}bingen, Auf der Morgenstelle 14, D-72076 T\"{u}bingen, Germany}

\author{Jean-Philippe Lansberg}
\email{lansberg@in2p3.fr}
\affiliation{IPNO, Universit\'e Paris-Sud, CNRS/IN2P3, F-91406, Orsay,  France}

\author{Cristian Pisano}
\email{c.pisano@nikhef.nl}
\affiliation{Nikhef and Department of Physics and Astronomy, VU University Amsterdam, De Boelelaan 1081, NL-1081 HV Amsterdam, The Netherlands}

\author{Marc Schlegel}
\email{marc.schlegel@uni-tuebingen.de}
\affiliation{Institute for Theoretical Physics,
                Universit\"{a}t T\"{u}bingen,
                Auf der Morgenstelle 14,
                D-72076 T\"{u}bingen, Germany}

\begin{abstract}
We argue that the study of heavy quarkonia, in particular that of $\Upsilon$, produced back to back 
with an isolated photon in $pp$ collisions at the LHC is the best --and currently unique-- way to access the 
distribution of both  the transverse momentum and the polarization of the gluon in an unpolarized proton. 
These encode fundamental information on the dynamics of QCD. We have derived analytical expressions for 
various transverse-momentum distributions which can  be measured at the LHC and which allow for a direct 
extraction of the aforementioned quantities. To assess the feasibility of such measurements, 
we have evaluated the expected yields and the relevant transverse-momentum distributions for different models 
of the gluon dynamics inside a proton.
\end{abstract}

\date{\today}

\maketitle

{\it Introduction.---} At LHC energies, the vast majority of hard reactions are initiated 
by the fusion of two gluons from both colliding protons. 
A good knowledge of gluon densities is therefore mandatory to perform reliable cross-section predictions, the
archetypal example being the $H^0$ boson production. In perturbative QCD (pQCD), the production cross section of a given particle
is conventionally obtained from the convolution of a hard parton-scattering amplitude squared and of the {\it collinear} 
parton distribution functions (PDFs) inside the colliding hadrons, $G(x,\mu)$ or $f_1^g(x,\mu)$  for the gluon~\cite{Brock:1993sz}. 
The PDF provides the distribution of a given parton in the proton as a function of its collinear 
momentum fraction $x$, at a certain (factorization) scale $\mu$. 
Whereas the scale evolution of the PDFs is given by pQCD, experimental data are necessary 
to determine their magnitude (see \eg~\cite{PDFfits}).

This {\it collinear} factorization, inspired by the parton model of Feynman and Bjorken, 
can be extended to take into account the transverse dynamics of the partons
inside the hadrons.  Different approaches have been proposed (unintegrated PDF, impact factors 
within $k_\sT$ factorization, etc.). Out of these, the Transverse-Momentum (TM) dependent factorization 
is certainly the most rigorous with proofs of factorization for a couple of processes~\cite{TMDfact,Ji:2004wu,Ma:2012hh,Zhu:2013yxa}.
The further advantage of the TM Dependent (TMD) formalism lies in its ability to deal with 
spin-dependent objects, both for 
the partons and the hadrons.

\begin{figure}[hbt!]
\centering
\subfloat[]{\includegraphics[width=0.25\columnwidth]{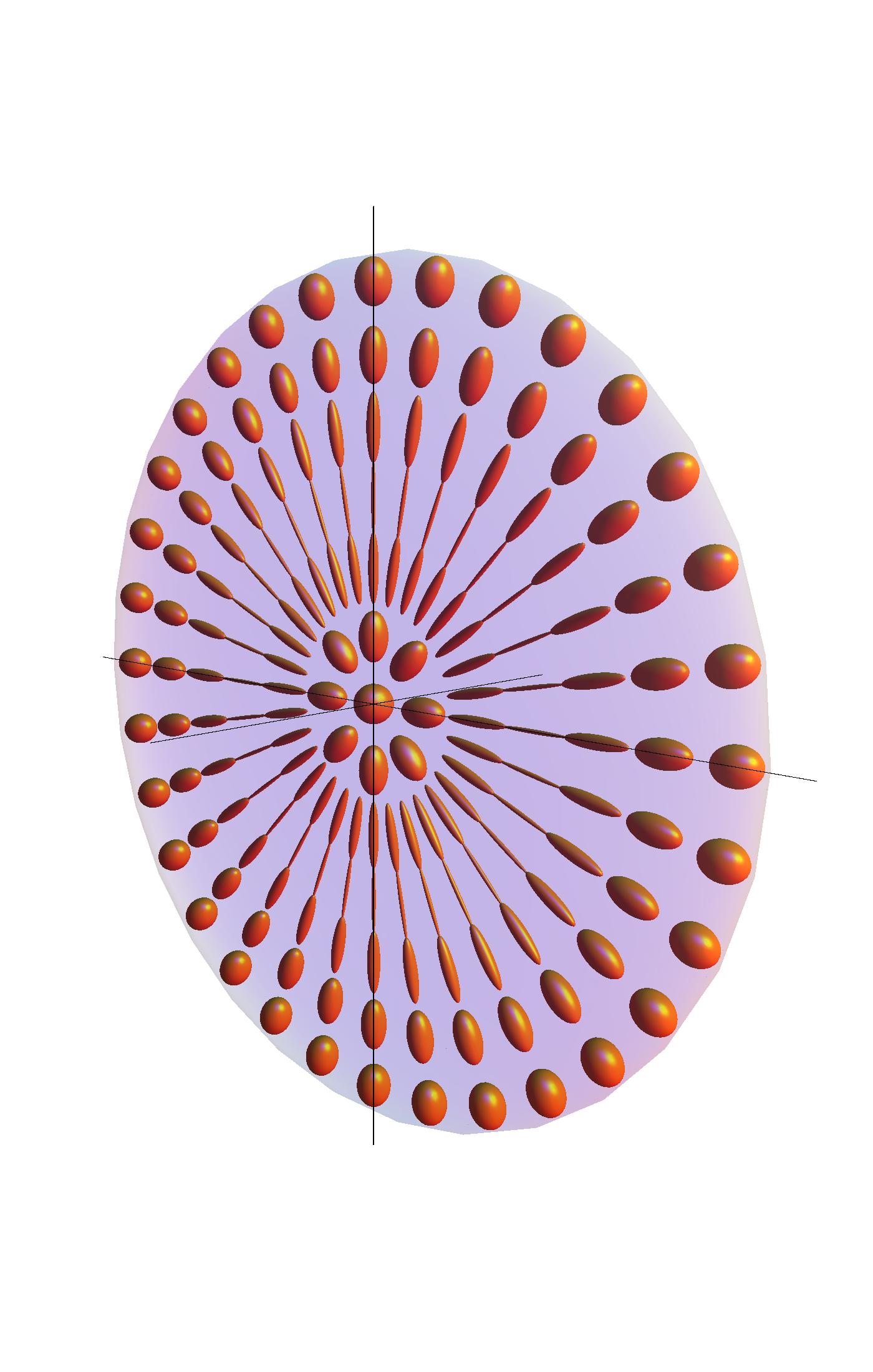}} \hspace*{-.2cm}
\subfloat[]{\includegraphics[width=0.25\columnwidth]{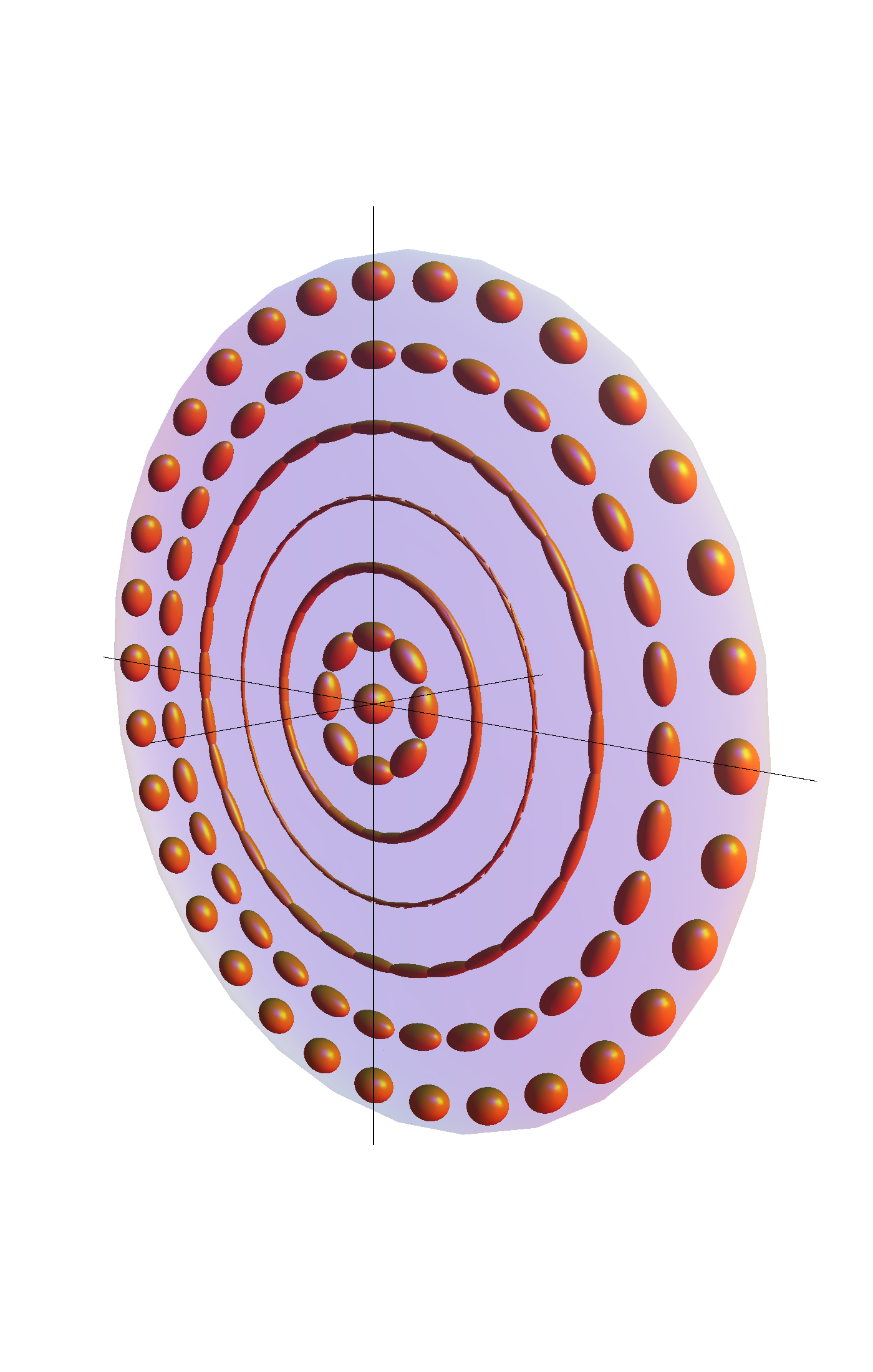}}
\subfloat[]{\includegraphics[width=0.5\columnwidth]{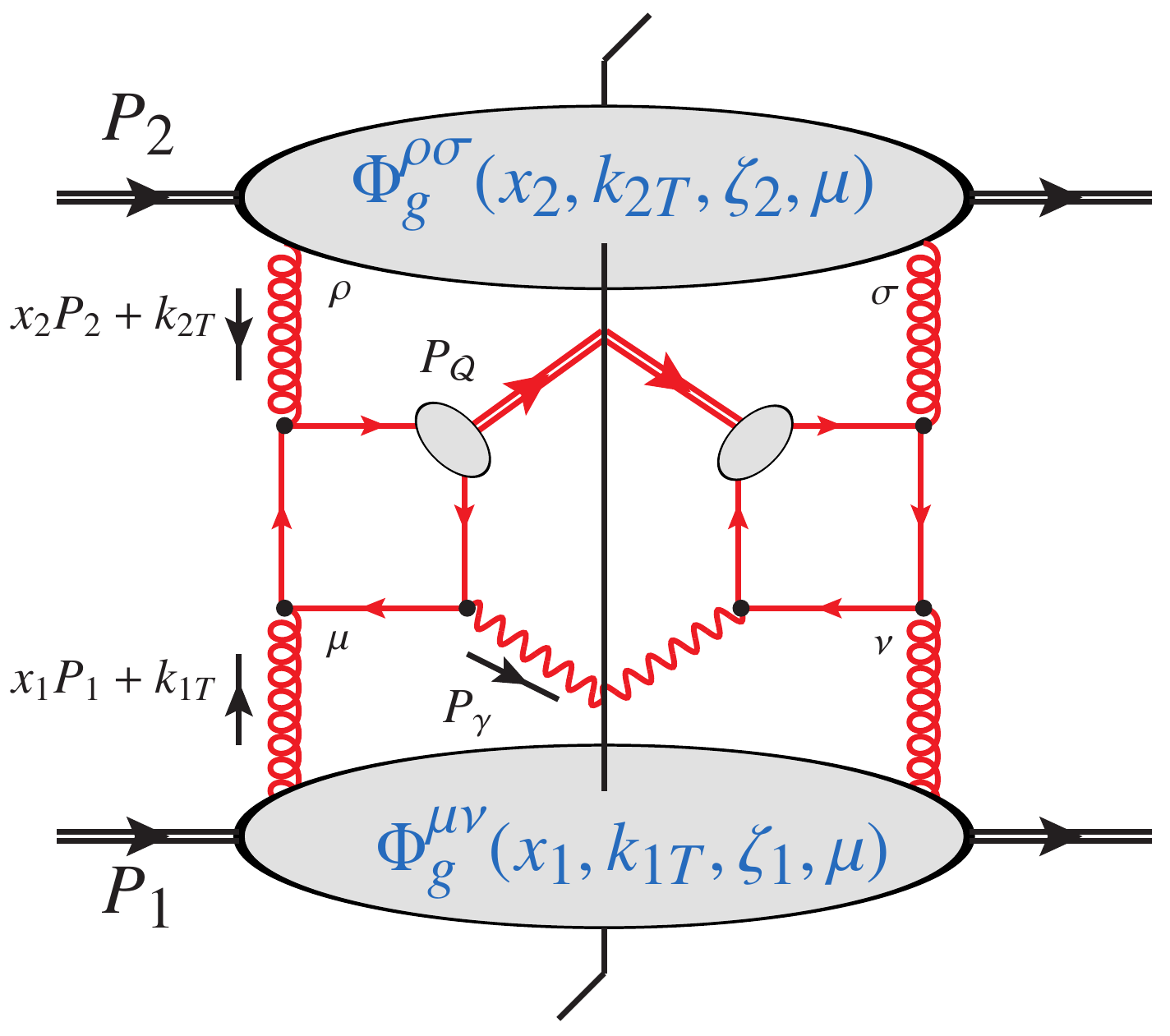}\label{Fig1:c}}
\caption{Visualization of the gluon polarization in the TM plane
for a positive (a) and negative (b) Gaussian $h_1^{\perp g}$.
[The ellipsoid major/minor axis lengths in the plane 
are proportional to the probability of finding a gluon with a linear polarization in that direction]. (c)
Feynman diagram for $p(P_1)\, {+}\, p(P_2) \to {\cal Q} (P_{\cal Q}) \,{+}  \,\gamma (P_\gamma) \,{+}X$ via 
gluon fusion at LO in the TMD-factorization formalism.
}
\label{Fig1}
\end{figure}

Much effort has been made recently to extract {\it quark} TMD distributions (TMDs in short) 
inside a proton from low energy data from HERMES, COMPASS or JLab experiments 
(see e.g.\ Ref.~\cite{QuarkTMD} for recent reviews). 
On the contrary, nothing is known experimentally about the {\it gluon} TMDs which rigorously parametrize the transverse motion of gluons inside a proton. For an unpolarized proton, these are  the distribution of unpolarized gluons, denoted by $f_1^g$, and the distribution of linearly-polarized gluons, $h_1^{\perp g}$~\cite{Mulders:2000sh}. These functions contain fundamental information on the transverse dynamics of the gluon content of the proton 
[see the interpretation of $h_1^{\perp g}$ in \cf{Fig1} (a-b)] and are necessary
to correctly describe gluon-fusion processes at all energies.
Without any knowledge of these functions it is impossible to calculate the 
Higgs transverse momentum distribution accurately \cite{TMDHiggs}.
We therefore stress that a first determination of these quantities should have high priority.

In the small-$x$ limit, the behavior of the  gluon TMD $f_1^g$ is probably connected to the Unintegrated Gluon Distribution (UGD)~\cite{smallxUGD}, which has been widely studied in the framework of the Color Glass Condensate (CGC) model \cite{Dominguez:2010xd,Dominguez:2011wm,Dominguez:2011br,Metz:2011wb}, in $k_\sT$-factorization approaches and as the solution of the CCFM equation \cite{Andersson:2002cf}. This connection is however less trivial than sometimes asserted, as in \eg~\cite{Hautmann:2013tba,Lipatov:2013yra}.  For instance, the Weizs\"acker-Williams distribution that appears in the CGC model {\it does} have the same operator structure as the TMD correlator (see \ce{eq:TMDcorrelator} below), {\it but} with a lightlike gauge link. The regularization of the rapidity divergence is thus different. Moreover, the CCFM equation does not rely on a gauge-invariant-operator definition. Nonetheless, to give some estimates of the experimental requirements, we will use various UGDs as an Ansatz for $f_1^g$ and let $h_1^{\perp\,g}$
saturate a model-independent positivity bound derived in Ref.~\cite{Mulders:2000sh}. The latter is in accordance with $k_\sT$-factorization in which full gluon polarization is implicit. In fact, this would serve as a test of the applicability of $k_\sT$-factorization methods for  $x$ close to  
$10^{-3}$.

In the following of this Letter, we argue that the LHC experiments are ideally positioned to
extract for the first time the gluon TMDs through the study of 
an isolated photon produced back to back with a heavy quarkonium. 
Furthermore, we show that the yields are large enough to perform such extractions with existing data at $\sqrt{s}=7$ and $8$~TeV.

{\it Reactions sensitive to  gluon TMDs.---} Several processes have been proposed to 
measure both $f_1^g$ and  $h_1^{\perp g}$. 
A potentially very clean probe to extract gluon TMDs is the back-to-back 
production of a heavy-quark pair in electron-proton collisions, 
$e\,p\to e \,Q \bar Q\,X$ in which the gluon TMDs appear  linearly. 
Theoretical predictions 
have been provided at leading order (LO)~\cite{ep2jetLO} and 
next-to-leading order (NLO)~\cite{Zhu:2013yxa} in pQCD. 
Such measurements could be performed at future facilities (EIC or LHeC), whose realization is however at best a decade away, while available HERA data on transverse momentum imbalance of dijets  (e.g., Ref.~\cite{HERAdata}) receive contributions from quark-induced subprocesses.

Back-to-back isolated photon-pair production in proton collisions, 
$p\,p\to \gamma\, \gamma \,X$ is also sensitive to gluon TMDs~\cite{Qiu:2011ai} 
and is accessible at RHIC and the LHC but suffers from a contamination from quark-induced channels, 
 a huge background from $\pi^0$-decays and an inherent difficulty to trigger on such events.

Final states such as a heavy-quark pair or a dijet~\cite{ep2jetLO} should also
be ideal candidates to probe gluon TMDs.
However, once there is a color flow into the detected final state in the 
partonic-scattering subprocess, one cannot cleanly separate final state interactions of this color flow from the non-perturbative TMD objects due to the non-Abelian characteristics of QCD~\cite{ColorEntanglement}. This leads to a breakdown of TMD factorization for processes with colored final states.

This problem can be avoided in the case of the production of heavy quarkonia, provided that the 
heavy-quark pair is produced in a colorless state at short distances as in the color-singlet model~\cite{CSM_hadron}, 
and that it is not accompanied by other --necessarily colorful-- partons. 
$C$-even quarkonium ($\chi_Q$, $\eta_Q$) production
at small TM  is one of these cases where the factorization
is expected to hold as illustrated by studies both at LO~\cite{ppJPsiLO} and NLO~\cite{ppJPsiNLO}. At low $P_{\mathcal{Q}T}$, 
$\eta_Q$ and $\chi_{Q0,2}$ production proceeds without the 
emission of a final-state gluon and the color-octet (CO)
contributions~\cite{NRQCD} are not kinematically enhanced. However, such experimental measurements are particularly difficult  
since they should be done at low TM, $P_{\mathcal{Q}T}\ll Q \simeq M_{\cal Q}$,  as required by TMD factorization.
The hard scale of the process, $Q$, can only be the mass of the heavy quarkonium, hence $Q \simeq M_{\cal Q}$.
The observation of low $P_{\mathcal{Q}T}$ $C$-even quarkonia is likely 
impossible with ATLAS and CMS. LHCb may look at these down to 
$P_{\mathcal{Q}T}\simeq 1$ GeV, but an unambiguous gluon-TMDs determination 
-- free of large power corrections in $P_{\mathcal{Q}T}/Q$ --
requires to reach the sub-GeV region. Besides, this would not allow one to look
 at the scale evolution of the TMDs. Only two ranges can be probed -- close to the charmonium and bottomonium masses. 

{\it Back-to-back  quarkonium+isolated-photon production.---}
We propose a novel process to overcome these issues : 
the production of a back-to-back pair of a $^3S_1$ quarkonium ${\cal Q}$ ($\Upsilon$ or $J/\psi$) 
and an isolated photon, $p\,p\to{\cal Q}+\gamma+X$. 
Compared to the aforementioned processes, it is accessible by the LHC experiments: only the TM
imbalance, $\bm{q}_\sT = \bm{P}_{{\cal Q} \sT} + \bm{P}_{\gamma \sT}$, has to be small, not the individual 
TM, for TMD factorization to apply. In addition, the hard scale of the process $Q$
 can be tuned by selecting different  invariant masses 
of the ${\cal Q}-\gamma$ pair. This allows one to look at the scale evolution of the TMDs
and to greatly increase the $q_T$-range  where the TMD factorization applies with tolerable
power corrections.

Previous studies~\cite{Kim:1996bb,Li:2008ym,Lansberg:2009db} have shown that 
the CO contributions to inclusive $\mc{Q}+\gamma$ production are likely smaller 
than in the inclusive case $\mc{Q}+X$ (see \eg~\cite{Brambilla:2010cs,Lansberg:2008gk,Lansberg:2006dh}) . [The case of $J/\psi+\gamma$ is however intriguing since
a state-of-the art NLO evaluations using recent NRQCD fits predict negative CO 
cross-sections~\cite{Li:2014ava}.] The smallness of CO contributions is crucial since these would violate the TMD factorization.

\begin{figure}[hbt!]
\centering
\includegraphics[width=0.9\columnwidth]{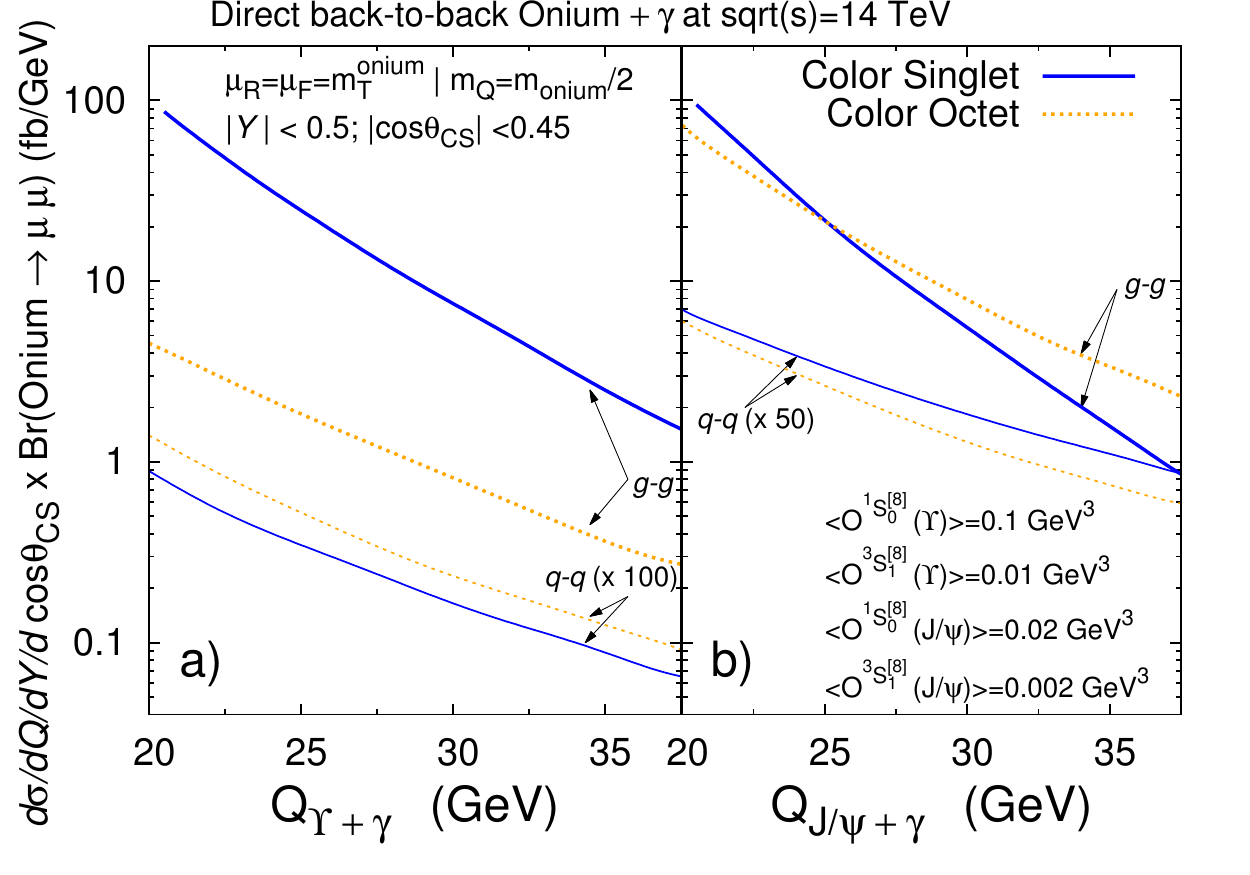}
\caption{Different contributions to the production of an isolated photon back to back with a) an $\Upsilon(1S)$ (resp. b) a $J/\psi$)  from $g-g$ and $q-\bar q$ fusion  from the CS and CO channels as function the invariant mass of the pair. The curves for the $q-\bar q$ fusion are rescaled by a factor 100 (resp. 50). The CO matrix elements we used are very close to those obtained in a recent LO fit of LHC data~\cite{Sharma:2012dy}.}
\label{fig:prod_channels}
\end{figure}

As studied in \cite{Mathews:1999ye}, the CO contributions are also suppressed w.r.t.\ the CS ones when the 
${\cal Q}-\gamma$ pair is produced back to back, \ie~dominantly from $2\to2$ processes, although the $gg$ fusion CS contribution (\cf{Fig1:c})
scales like $P_{\mathcal{Q}T}^{-8}$. Indeed, the $P_{\mathcal{Q}T}^{-4}$ (fragmentation) CO contribution only appears
for $q\bar q$ annihilation  --extremely suppressed at LHC energies-- and, incidentally, on the order
of  the pure QED CSM contribution (as for $J/\psi+W$~\cite{Lansberg:2013wva}). As regards $gg$ fusion
CO channels, they are subleading in $P_{\mathcal{Q}T}$, since they come from quark box and $s$-channel gluon diagrams, only via
$C=+1$ CO states, such as $^1S_0^{[8]}$ or $^3P_J^{[8]}$.
[For the $J/\psi$, these CO states are known to 
be severely constrained if one wants to comply with $e^+e^-$ inclusive data~\cite{Zhang:2009ym}.] To substantiate this, 
we have computed the different CS and CO contributions in LO NRQCD,
see \cf{fig:prod_channels}. The CS yield is clearly dominant for 
the $\Upsilon$ and likely above the CO one for the $J/\psi$ at the lowest $Q$ accessible at the LHC ($P_{\mathcal{Q}T} \gtrsim 10$ GeV).
It is also clear that this process is purely from $gg$ fusion.

A further suppression of CO contributions can be achieved by also isolating the 
quarkonium (see~\cite{Kraan:2008hb}). The isolation 
should be efficient at large enough $P_{\mathcal{Q}T}$ where the soft partons emitted during 
the hadronization of the CO heavy-quark pair are
boosted and energetic enough to be detected. Experimentally, this would provide an 
interesting check of the CS dominance by measuring
the (conventional) $q_T$-integrated cross section which should coincide with the parameter-free  CSM 
prediction. This would also confirm that double-parton scattering contributions
are suppressed by the isolation criteria. We emphasize that, according to our evaluations, 
such an isolation is not at all necessary for the $\Upsilon$ case.

{\it Analytical expression for the $q_T$-dependent cross section.---}
Within TMD factorization (\cf{Fig1:c}), the cross section for a gluon-fusion inititiated process
is written, up to $\mc{O}(\qT^2/Q^2)$ corrections, 
as the convolution of a hard part with two TM dependent correlators, \ie~
\begin{multline}\label{eq:factformula}
\d\sigma = \frac{(2\pi)^4}{8 s^2}\!
  \int\!\! \d^{2}\koneT \d^{2}\ktwoT
  \delta^{2}(\koneT + \ktwoT - \qT)
  \mc{M}_{\mu\rho}
  \left(\mc{M}_{\nu\sigma}\right)^*\\
  \Phi_g^{\mu\nu}(x_1,\koneT,\zeta_1,\mu)\,
  \Phi_g^{\rho\sigma}(x_2,\ktwoT,\zeta_2,\mu) \d\mc{R},
\end{multline}
where $s = (P_1 + P_2)^2$ is the hadronic center-of-mass system (c.m.s.) energy squared
and the phase space element of the outgoing particles is denoted by $\d\mc{R}$.
The hard part can be obtained as a series expansion in $\alpha_s$ by perturbatively
calculating the partonic scattering $g(k_1) \,{+} \,g(k_2) \to  {\cal Q}(P_{\cal Q})\, {+} \, \gamma(P_\gamma)$,
with the incoming gluon momenta given by $k_1 = x_1 P_1 + k_{1 \sT} - k_{1 \sT}^2/(x_1 s) P_2$ (and likewise for $k_2$),
and subtracting the parts already contained in the gluon TMD correlators~\cite{Ji:2005nu,Sun:2011iw,Ma:2012hh}.
$k_{1\sT}$ is a 4-vector perpendicular to both $P_1$ and $P_2$,
which has transverse components $\koneT$ in the c.m.s. frame; $x_1={q{\cdot} P_2}/{P_1{\cdot} P_2}$ 
and $x_2 = {q{\cdot} P_1}/{P_1{\cdot} P_2}$, where $q = P_\mc{Q} + P_\gamma$.

Since QCD corrections to the inclusive production of a quarkonium-photon pair are known 
to be large~\cite{Li:2008ym,Lansberg:2009db}, we find it useful to emphasize that this does not translate to TMD factorization. 
The reason is that the initial-state radiations are absorbed into the TMDs  such that the hard part is 
free of $q_\sT$-dependence and, with appropriate choices of $\zeta$ and $\mu$, is also free of 
large logarithms~\cite{Ji:2005nu,Sun:2011iw,Ma:2012hh}. 
In addition, the back-to-back (small $q_\sT$) requirement and the photon isolation in our observable further 
suppresses additional radiations. 
A LO calculation of the hard part is therefore sufficient for a first gluon TMD extraction.

The gluon-TMD correlator for an unpolarized proton is defined as
\begin{multline}\label{eq:TMDcorrelator}
\Phi_g^{\mu\nu}(x,\kT,\zeta,\mu) \equiv
	      \int \frac{\d(\xi{\cdot} P)\, \d^2 \xi_\sT}{(x P{\cdot} n)^2 (2\pi)^3}\,
	      e^{i ( xP + k_\sT) \cdot \xi}\times\\
	      \qquad\qquad\qquad\qquad\langle P| F_a^{n\nu}(0)
	      \left(\mc{U}_{[0,\xi]}^{n[\text{--}]}\right)_{ab} F_b^{n\mu}(\xi)
	      |P\rangle \Big|_{\xi \cdot P^\prime = 0}\\
=	-\frac{1}{2x} \bigg \{g_\sT^{\mu\nu} f_1^g
	-\bigg(\frac{k_\sT^\mu k_\sT^\nu}{M_p^2}\,
	{+}\,g_\sT^{\mu\nu}\frac{\kT^2}{2M_p^2}\bigg)\,
	h_1^{\perp\,g} \bigg \} + \text{suppr.},
\end{multline}
where $g^{\mu\nu}_{\sT} = g^{\mu\nu} - (P_1^{\mu}P_2^{\nu}+P_2^\mu P_1^\nu)/P_1{\cdot}P_2$, $M_p$ is the proton mass and 
the gauge link 
$\mc{U}_{[0,\xi]}^{n[\text{--}]}$ renders the matrix element gauge invariant. It runs from $0$ to $\xi$ 
via $-\infty$ along the $n$ direction. [$n$ is a timelike 
dimensionless 4-vector with no transverse components such that
$\zeta^2 = (2n{\cdot}P)^2/n^2$.] 
The correlator is parametrized by the two gluon TMDs 
discussed above, $f_1^g(x,\kT,\zeta,\mu)$ and $h_1^{\perp\,g}(x,\kT,\zeta,\mu)$ \cite{Mulders:2000sh}
and by terms that are suppressed in the high-energy limit.

The  structure of the TMD cross section is then found to be
\begin{align}\label{eq:crosssection}
&\frac{\d\sigma}{\d Q \d Y \d^2 \qT \d \Omega} 
  =\frac{C_0(Q^2 - M_\mc{Q}^2)}{s\, Q^3 D}\Bigg\{
  F_1\, \mc{C} \Big[f_1^gf_1^g\Big]+ F_3\cos(2\phi) \nonumber\\  &\mc{C} \Big[w_3 f_1^g h_1^{\perp g} + x_1\! \leftrightarrow\! x_2 \Big]\!+\!F_4\! \cos(4\phi\!)\, \mc{C}\! \left[w_4 h_1^{\perp g}h_1^{\perp g}\right]\!\Bigg\} 
  + \mc{O}\! \left(\frac{\qT^2}{Q^2} \right),
\end{align}
where $\d\Omega=\d\!\cos\theta\d\phi$ is expressed in terms of Collins-Soper angles \cite{Collins:1977} and where 
$Q$, $Y$ and $\qT$ are the invariant mass, the rapidity and the TM of the pair --the latter two to be measured in the hadron c.m.s. frame. The Collins-Soper angles describe the spatial orientation of the back-to-back photon-quarkonium pair in the Collins-Soper rest frame of the pair. The overall normalization 
is given by $C_0= 4 \alpha_s^2 \alpha_{{em}} e_Q^2 |R_0(0)|^2/(3M_\mc{Q}^3)$, where $R_0(0)$ is the quarkonium radial wave function at the origin and $e_Q$ the  heavy quark charge. 
The $F$ factors, the denominator $D$ and the weights are found to be
\begin{align}
 F_1 &=  1 + 2 \alpha ^2 + 9 \alpha ^4 + (6 \alpha ^4-2) \cos^2\theta + (\alpha ^2-1)^2 \cos^4\theta ,
 \nonumber\\
 F_3 &=  4\, \alpha^2\, \sin^2\theta,\ \  F_4 =  (\alpha ^2-1)^2 \sin^4\theta  ,\nonumber\\
 D &= \left((\alpha ^2+1)^2-(\alpha ^2-1)^2 \cos^2 \theta\right)^2,\nonumber\\
 w_3 		&= \frac{\qT^2\ktwoT^2 - 2 (\qT{\cdot}\ktwoT)^2}{2 M_p^2 \qT^2},\nonumber\\
 w_4		&=  2\left[\frac{\koneT{\cdot}\ktwoT}{2M_p^2} - 
		\frac{(\koneT{\cdot}\qT) (\ktwoT{\cdot}\qT)}{M_p^2\qT^2}\right]^2 -\frac{\koneT^2\ktwoT^2 }{4 M_p^4}.
\end{align}
where $\alpha \equiv Q/M_\mc{Q}$.
The convolution is defined as
\begin{multline}
\mathcal{C}[w\, f\, g] \equiv \int\!\! \d^{2}\koneT\!\! \int\!\! \d^{2}\ktwoT\,
  \delta^{2}(\koneT+\ktwoT-\bm q_{\sT})\times\\
  w(\koneT,\ktwoT)\, f(x_1,\koneT^{2})\, g(x_2,\ktwoT^{2}),
\end{multline}
where $x_{1,2} =  \exp[\pm Y]\, Q/\sqrt{s}$.

\begin{figure*}[htb]
\centering
\subfloat[]{\includegraphics[width=0.33\textwidth]{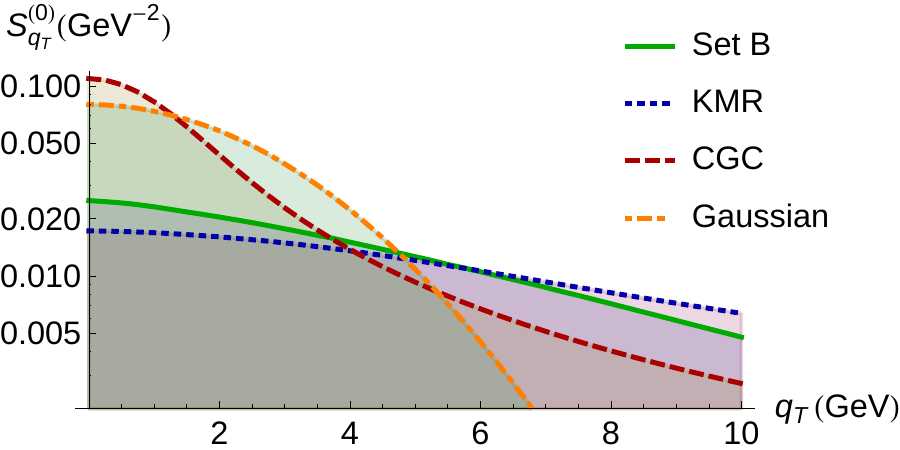}\label{fig:dsigma0dqT}}
\subfloat[]{\includegraphics[width=0.33\textwidth]{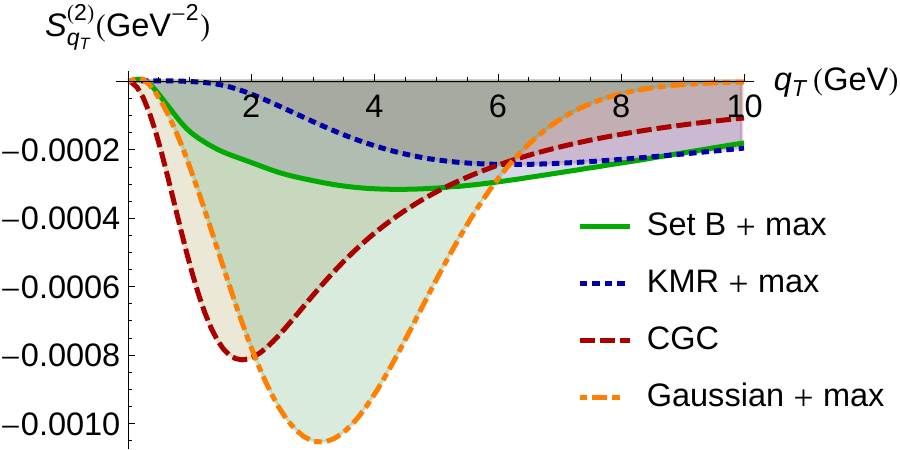}\label{fig:dsigma2dqT}}
\subfloat[]{\includegraphics[width=0.33\textwidth]{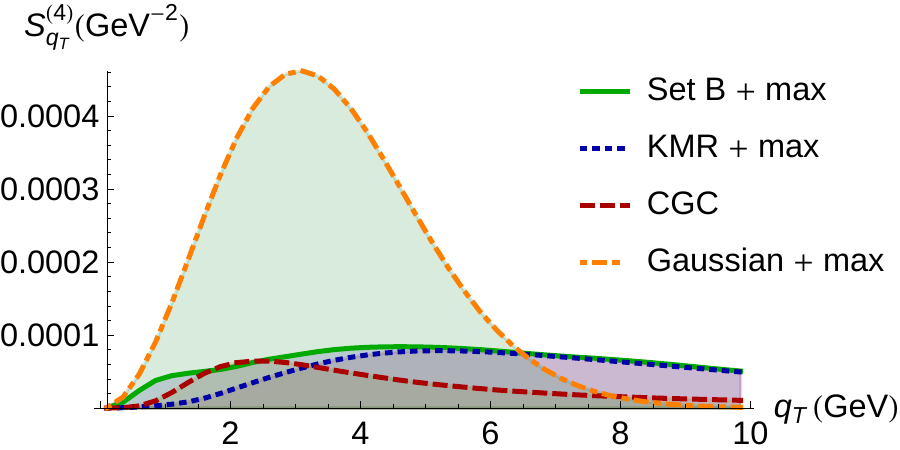}\label{fig:dsigma4dqT}}
\caption{Model predictions for $\Upsilon+\gamma$ production at $Q=20$ GeV, $Y=0$  and $\theta =\pi/2$ at $\sqrt{s}=14$ TeV
for (a) ${\cal S}^{(0)}_{q_T}$, (b) ${\cal S}^{(2)}_{q_T}$ and
(c) ${\cal S}^{(4)}_{q_T}$. The longitudinal momentum fractions are $x_1=x_2=Q/\sqrt{s}\simeq1.4\times10^{-3}$.}
\end{figure*}

We propose the measurement of 3 TM spectra, normalized and weighted by
$\cos n\phi$ for $n=0,2,4$:
\begin{equation}
{\cal S}^{(n)}_{q_T}\!\equiv \! \frac{\int \!\!\d\phi \cos(n\, \phi\!)\, \frac{\d\sigma}{\d Q \d Y \d^2 \bm q_\sT \d \Omega}}
{\int\!\! \d \bm q_\sT^2 \int \!\!\d\phi \frac{\d\sigma}{\d Q \d Y \d^2 \bm q_\sT \d \Omega}},
\end{equation}
where we will take the $\bm q_\sT^2$ integration in the denominator up to $(Q/2)^2$.
These spectra separate out the the 3 terms in \ce{eq:crosssection}: 
\begin{align}\label{eq:qTdistrs}
{\cal S}^{(0)}_{q_T}\!&=\!\frac{\mc{C}[f_1^g f_1^g]}
  {\int\!\! \d \bm q_\sT^2\, \mc{C}[f_1^g f_1^g]},
  {\cal S}^{(2)}_{q_T}= 
  \frac{F_3\, \mc{C}[w_3 f_1^g h_1^{\perp g} + x_1 \leftrightarrow x_2]}
  {2 F_1\! \int\!\! \d \bm q_\sT^2\, \mc{C}[f_1^g f_1^g]},\nonumber\\
{\cal S}^{(4)}_{q_T}\!&=  
 \frac{F_4\, \mc{C}[w_4 h_1^{\perp g} h_1^{\perp g}]}
  {2 F_1\! \int\!\! \d \bm q_\sT^2\, \mc{C}[f_1^g f_1^g]}.
\end{align}

It is remarkable to note that the sole measurement of ${\cal S}^{(0)}_{q_T}$, 
\ie~of the cross section integrated over 
  $\phi$,
 allows for a clean determination of the 
unpolarized gluon TMD, $f_1^g$, since $h_1^{\perp g}$ does not enter ${\cal S}^{(0)}_{q_T}$. 
If ${\cal S}^{(2)}_{q_T}$ or ${\cal S}^{(4)}_{q_T}$ can also be measured, 
then the linearly-polarized gluon distribution, $h_1^{\perp g}$, is also accessible.

{\it Numerical results and discussions.---} In our calculations we adopt the following 
UGD Ans\"atze for $f_1^g$: the {\it Set B0} solution to the CCFM equation with an initial
distribution based on the HERA data from \cite{Jung:2004gs,Jung:2010si},
the KMR parametrization from \cite{Kimber:2001sc} and the CGC
model prediction from \cite{Dominguez:2010xd,Dominguez:2011wm,Dominguez:2011br,Metz:2011wb}.
The first two depend on a factorization scale, taken to be  $Q$, whereas
the last one depends on a saturation scale taken as 
$Q_s= (x_0/ x)^\lambda Q_0$,
with $\lambda=0.29$, $x_0 = 4\cdot 10^{-4}$ and $Q_0 = 1$ GeV~\cite{Gelis:2010nm}.
We have also used a simple Gaussian parametrization, 
as done in~\cite{Sridhar:1998rt} to describe
the intrinsic gluon TM, but with  $\kTsqav= (2.5\text{ GeV})^2$.
Our results are shown in Fig.\ \ref{fig:dsigma0dqT}.
 
 For $h_1^{\perp g}$, we use the CGC model prediction of~\cite{Dominguez:2011br,Metz:2011wb} and the maximal value from the positivity constraint  $\vert h_1^{\perp g}\vert \leq 2M_p^2/\koneT^2 f_1^g$~\cite{Mulders:2000sh}. 
The resulting ${\cal S}^{(2,4)}_{q_T}$  are plotted in \cf{fig:dsigma2dqT}
and \cf{fig:dsigma4dqT}.
 
From \cf{fig:dsigma0dqT}, we first conclude that measuring ${\cal S}^{(0)}_{q_T}$
in bins of 1 GeV should suffice to get a first determination of the shape of the
unpolarized gluon distribution.
As regards ${\cal S}^{(2)}_{q_T}$ and ${\cal S}^{(4)}_{q_T}$, whose magnitude is obviously smaller, 
one can integrate them over $\bm q_\sT^2$  (up to $(Q/2)^2$) to get the {\it first experimental verification} of a 
nonzero linearly-polarized gluon distribution.
${\cal S}^{(2)}_{q_T}$ is here the most promising as we obtain for the integrated
distribution $-2.9\%$, $-2.6\%$, $-2.5\%$ and $-2.0\%$ for the Gauss, CGC, SetB and KMR Ansatz respectively,
whereas for the $n=4$ distribution we obtain $1.2\%$, $0.7\%$, $0.6\%$, and $0.3\%$
for the Gauss, SetB, KMR and CGC model respectively. We note that the $q_\sT$-integrated cross section for $\Upsilon+\gamma$ production in Fig.~\ref{fig:prod_channels} is about 100 (50) fb/GeV at $Q=20$ GeV for $\sqrt{s}= 14 (7)$ TeV.
The 20 fb$^{-1}$ of integrated luminosity already collected at $7+8$ TeV should be sufficient to measure the $q_\sT$ shape of $S^{(0)}_{q_\sT}$, while
$S^{(2)}_{q_\sT}$ could be measured in a single $q_\sT$-bin.

{\it Conclusion.---}
The production of an isolated photon back to back with a --possibly isolated-- quarkonium 
in $pp$ collisions is the ideal observable to study the transverse dynamics
and the polarization of the gluons in the proton along the lines of TMD factorization.
The requirement for a heavy quarkonium in the final state suppresses quark-initiated reactions
making it a very clean probe of the gluon content of the proton, 
whereas the large scale set by the invariant mass of the pair allows 
a TMD-factorized description over an extensive range of $q_\sT$ and hence
an extraction of the gluon TMDs in this range.
The expected yields at the LHC experiments are large enough to get the first experimental verification of a nonzero gluon polarization in unpolarized protons.
These measurements would therefore provide a test of the reliability of the
$k_\sT$-factorization approach at $x\sim 10^{-3}$ and allow for the first extraction of the gluon TMDs in the proton.

\begin{acknowledgments}\small
We thank D.\ d'Enterria, V.\ Kartvelishvili, C.\ Lorc\'e, A.\ Signori, L.\ Szymanowski, S.\ Wallon and J.X.\ Wang for useful discussions. This work was supported in part by the French CNRS, grants
PICS-06149 Torino-IPNO \& PEPS4AFTER2, by the German Bundesministerium f\"{u}r Bildung und Forschung (BMBF),
grant no. 05P12VTCTG, by the European Community under the “Ideas” program QWORK (contract 320389), and by the Marie Curie grant IRG 256574.
\end{acknowledgments}

\end{document}